\documentstyle[manuscript,aps]{revtex}
\tightenlines
\begin{document}

\title{Application of the Exact Muffin-Tin Orbitals Theory:
the Spherical Cell Approximation}

\author{L. Vitos$^1$, H. L. Skriver$^2$, B. Johansson$^1$,
J. Koll\'ar$^3$}

\address{$^1$Condensed Matter Theory Group, Physics Department, \\
     Uppsala University, S-75121 Uppsala, Sweden}

\address{$^2$Center for Atomic-scale Materials Physics
and Department of Physics,\\
Technical University of Denmark, DK-2800 Lyngby, Denmark}

\address{$^3$Research Institute for Solid State Physics, \\
H-1525 Budapest, P.O.Box 49, Hungary}

\date{8 October 1999}

\maketitle

\begin{abstract}

We present a self-consistent electronic structure calculation method based 
on the {\it Exact Muffin-Tin Orbitals} (EMTO) Theory developed by O. K. 
Andersen, O. Jepsen and G. Krier (in {\it Lectures on Methods of Electronic 
Structure Calculations}, Ed. by V. Kumar, O.K. Andersen, A.  Mookerjee, Word 
Scientific, 1994 pp. 63-124) and O. K. Andersen, C. Arcangeli, R. W. Tank, T. 
Saha-Dasgupta, G. Krier, O. Jepsen, and I. Dasgupta,
(in {\it Mat. Res. Soc. Symp. Proc.} {\bf 491}, 1998 pp. 3-34).
The EMTO Theory can be considered as an 
{\it improved screened} KKR (Korringa-Kohn-Rostoker) method which is able to 
treat large overlapping 
potential spheres. Within the present implementation of the EMTO Theory the one 
electron equations are solved exactly using the Green's function
formalism, and the Poisson's equation is solved within the {\it Spherical Cell
Approximation} (SCA). To demonstrate the accuracy of the SCA-EMTO method
test calculations have been carried out.
\end{abstract}
\vspace{10mm}

\narrowtext
\pagebreak

\section{Introduction}

During the last decades many attempts have been made to develop 
accurate and at the same time efficient methods for solving the 
Kohn-Sham equations in an application of the Density Functional Theory 
for condensed matter. The accuracy of the methods is crucial e.g. when one 
searches for the answers given by different density functional 
approximations. The full-potential techniques have been specially designed 
to fulfill this requirement. Though, in principle, they give highly 
accurate results, they have their own limitations. In many cases a compromise 
has been made between the accuracy and efficiency, and methods based on 
approximate one electron potentials have been developed. The most commonly 
used muffin tin approach,
albeit its mathematical formulation is very elegant, presents a rather poor 
representation of the exact potential. Though, the {\it Atomic Sphere 
Approximation} (ASA) brings a real improvement to the potential, most of the 
conventional methods based on the ASA use similar approximation to the one 
electron energies and charge density as well \cite{asato99}. 
Therefore, using these methods, reasonably accurate 
results can only be obtained for close packed systems, and
they are not suitable to treat systems of low symmetry. In order to 
maintain or increase the accuracy different corrections should be included 
and, therefore, the ASA based methods lose their elegance and efficiency.

A few years ago breakthrough was made by developing the Exact 
Muffin-Tin Orbitals (EMTO) Theory \cite{andersen94,third}. Within the EMTO 
Theory 
the one electron states are calculated exactly for the overlapping muffin-tin 
potential, while the solution of Poisson's equation can 
include certain shape approximations, if required. By separating the two 
approaches used 
for the potential and one electron states the accuracy can be sustained at a 
level comparable with that of the full-potential techniques without losing 
significantly from the efficiency. The EMTO Theory can be considered as an 
improved screened KKR method \cite{andersen92,szunyog94}, within that large 
overlapping potential spheres can be used for accurate representation of the 
exact one electron potential \cite{third,arcangeli}. 

In this work we present a self-consistent implementation of the EMTO Theory 
within the {\it Spherical Cell Approximation} (SCA) for the Poisson's equation. 
In the first part we review the EMTO Theory \cite{andersen94,third}, the 
definition 
of the screened spherical waves and the matching 
equation. Furthermore, we establish the expressions 
for the number of states and electron density using the Green's function 
formalism. In the second part of the paper we discuss the impact of the SCA, 
used for the shape of the Wigner-Seitz cell, on the total charge density 
and on the overlapping muffin tin potential. Finally, we establish the 
accuracy of the SCA-EMTO method by performing test calculations for some 
systems where reliable full-potential data are available.
An approximate solution of the kink cancellation equation in order 
to reduce the number of iterations needed in a self-consistent calculation is 
presented in the Appendix.

\section{Overview: the EMTO Theory}

In the following we review the basic concepts of the EMTO Theory developed 
by O. K. Andersen and co-workers \cite{andersen94,third,andersen96}. 
Assume that the one-electron Kohn-Sham equations,

\begin{equation}\label{scheq}
\Bigl [ -\nabla^2\;+\;v({\bf r})\Bigr ]\;\Psi_j({\bf r})\;=\;\epsilon_j\;
\Psi_j({\bf r}),
\end{equation}
are solved within the muffin tin approximation for the effective potential,

\begin{equation}\label{mtpot}
v({\bf r})\;\approx\;v_0\;+\;\sum_R\;\left[ v_R(r_R)\;-\;v_0\right],
\end{equation}
where $R$ runs over the lattice sites. Here and in the following we use 
the notation ${\bf r}_R\equiv{\bf r} - {\bf R}$ and omit the vector notation 
for the index $R$. In (\ref{mtpot}) $v_0$ denotes a parameter \cite{andersen96}
that reduces to the muffin tin zero in the case of non-overlapping muffin tins.
The spherical potentials $v_R(r_R)$ become equal to $v_0$ outside the potential
spheres of radii $s_R$. These radii can be chosen as the linear overlap between
the spheres to be as large as $30-40\;\%$ \cite{andersen92,third,arcangeli}.
It has turned out that for a good representation 
of the real full-potential in terms of overlapping muffin tin wells
usually a big overlap is preferred between the potential spheres 
\cite{arcangeli}.
 
In order to solve the Schr\"odinger equation (\ref{scheq}) for the 
muffin tin potential (\ref{mtpot}) one chooses different basis functions 
inside the potential spheres and in the interstitial region.
Inside the sphere at {\bf R} the partial waves are 
chosen as the basis function which are defined as the products of the 
regular solutions of the radial Schr\"odinger equation,

\begin{equation}\label{radsch}
\frac{\partial^2 \Bigl[r_R\;\phi_{Rl}(\epsilon,r_R)\Bigr]}{{\partial r_R}^2}
\;=\;\Bigl[\frac{l(l+1)}{r^2_R}+v_R(r_R)-\epsilon\Bigr]\;r_R\;
\phi_{Rl}(\epsilon,r_R),
\end{equation}
and the real spherical harmonics, viz.

\begin{equation}
\phi_{RL}(\epsilon,{\bf r}_R)\;=\;\phi_{Rl}(\epsilon,r_R)\;Y_L({\hat r}_R),
\end{equation}
where $L=(l,m)$. Outside the spheres the so called {\it screened 
spherical waves}, $\psi^a_{RL}(\kappa,{\bf r}_R)$, are used as basis functions.
Therefore, the wave function for the energy $\epsilon_j$ can be written as

\begin{equation}\label{wavefc}
\Psi_j({\bf r})\;=\;\sum_{RL}\;
\phi_{RL}(\epsilon_j,{\bf r}_R)\;\Theta_R({\bf r}_R)\;u^a_{RL,j}\;+\;
                    \sum_{RL}\;
\psi^a_{RL}(\kappa_j,{\bf r}_R)\;[1-\Theta_R({\bf r}_R)]\;v^a_{RL,j}.
\end{equation}
Here $\kappa^2= \epsilon -v_0$, and in the non-overlapping muffin tins limit
it denotes the interstitial one-electron kinetic energy. The 
$\Theta_R({\bf r}_R)$ is one inside the sphere of radius $s_R$
centered at {\bf R} and zero outside. The expansion coefficients $u^a_{RL,j}$ 
and $v^a_{RL,j}$ as well as the 
energies $\epsilon_j$ are determined from the condition that the wave function
$\Psi_j({\bf r})$ and its first derivative should be continuous
at the potential spheres. The algebraic formulation of this matching condition 
in the EMTO formalism is the so the called {\it kink cancellation
equation}, which is equivalent to the KKR 
(Korringa-Kohn-Rostoker) equation in an arbitrarily screened representation
\cite{szunyog94}.

\subsection{The screened spherical waves}

The screened spherical waves can be defined \cite{andersen94} in conjunction 
with hard spheres
centered at all sites ${\bf R}$ with radii $a_{Rl}$. They are solutions of the 
wave equation,

\begin{equation}
\Bigl [\nabla^2\;+\kappa^2\Bigr ]\;\psi^a_{RL}(\kappa^2,{\bf r}_R)\;=\;0,
\end{equation}
with the boundary condition that on their own $a-$spheres they behave 
like a pure real spherical harmonic, while the 
$Y_{L'}({\hat r}_{R'})$ projections on all the other $a-$spheres, $R'\ne R$, 
vanish. They form a complete basis set in the "$a$" interstitial region and 
may
be expressed in terms of the "value", $f^a_{RL}$, and the "slope", $g^a_{RL}$, 
functions \cite{andersen94}, whose radial part satisfy the following boundary 
conditions

\begin{eqnarray}
f^a_{Rl}(\kappa,r)\vert_{a_{Rl}}\;=\;1\;\;\mbox{and}\;\;
\frac{\partial f^a_{Rl}(\kappa,r)}{\partial r}\vert_{a_{Rl}}\;=\;0,\\
g^a_{Rl}(\kappa,r)\vert_{a_{Rl}}\;=\;0\;\;\mbox{and}\;\;
\frac{\partial g^a_{Rl}(\kappa,r)}{\partial r}\vert_{a_{Rl}}\;=\;\frac{1}
{a_{Rl}}.
\end{eqnarray}
These functions may of course be expressed in terms of the usual spherical
Bessel and Neumann functions \cite{abramovitz}

\begin{equation}
f^a_{Rl}(\kappa,r) = j_l(\kappa r){\mathcal W}_a\{f,\kappa\;n\}\;-
\;\kappa\;n_l(\kappa r){\mathcal W}_a\{f,j\}
\end{equation}
and
\begin{equation}
g^a_{Rl}(\kappa,r) = j_l(\kappa r){\mathcal W}_a\{g,\kappa\;n\}\;-
\;\kappa\;n_l(\kappa r){\mathcal W}_a\{g,j\}
\end{equation}
since ${\mathcal W}_r\{j,n\}=1/\kappa$,
and therefore satisfy the Wronskian

\begin{equation}
{\mathcal W}_r\{f^a_{Rl},g^a_{Rl}\}\;\equiv\;r^2\;\Bigl[f^a_l(\kappa,r)\;
\frac{\partial g^a_l(\kappa,r)}{\partial r}\;- \;
\frac{\partial f^a_l(\kappa,r)}{\partial r}\;
g^a_l(\kappa,r)\Bigr]\;=\;a_{Rl}.
\end{equation}
The screened spherical wave $\psi^a_{RL}(\kappa^2,{\bf r}_R)$ may be
expanded in spherical harmonics $Y_{L'}({\hat r}_{R'})$ about any site
${\bf R'}$, as

\begin{equation}\label{ssw}
\psi^a_{RL}(\kappa^2,{\bf r}_R)\;=\;f^a_{Rl}(\kappa,r_R)\;Y_L({\hat r}_R)
\delta_{RR'}\;\delta_{LL'}
\;+\; \sum_{L'}g^a_{R'l'}(\kappa,r_{R'})\;Y_{L'}({\hat r}_{R'})\;
S^a_{R'L'RL}(\kappa^2),
\end{equation}
where the expansion coefficients, $S^a_{R'L'RL}(\kappa^2)$, are the elements
of the so called {\it slope matrix}. The slope matrix can be derived from the 
bare KKR structure constant matrix $B_{R'L',RL}(\kappa)$, by matrix inversion
\cite{andersen94}

\begin{equation}\label{dyson}
S^a(\kappa^2) = {\mathcal D}\{j(\kappa,a)\}-
\frac{1}{a\;j(\kappa,a)} \Bigl[-B(\kappa) \;+\;\kappa\; cot\;\alpha(\kappa)
\Bigr]^{-1}\frac{1}{j(\kappa,a)},
\end{equation}
where ${\mathcal D}$ denotes the logarithmic derivative, 
${\mathcal D}\{j(r)\}\equiv r \left[\partial j(r)/\partial r\right]/j(r)$, 
and for simplicity where we have used matrix notation. We note that 
this equation is equivalent to
Eq.\ (3.26) from Ref.\ \cite{andersen94} and Eq.\ (15) from Ref.\ \cite{third}.
The bare KKR structure constants are defined as the expansion coefficients 
of the $\kappa\;n_L(\kappa,{\bf r}_R)\;\equiv\;\kappa\;n_l(\kappa r_R)\;
Y_L({\hat r}_R)$ functions around site $R'$ in terms of the 
$j_{L'}(\kappa,{\bf r}_{R'})\;\equiv\;j_{l'}(\kappa r_{R'})\;
Y_{L'}({\hat r}_{R'})$ functions, i.e.

\[
\kappa\;n_L(\kappa,{\bf r}_R)\;=\;\sum_{L'} j_{L'}(\kappa,{\bf r}_{R'})\;
B_{R'L'RL}(\kappa),
\]
with
\begin{equation}
B_{R'L'RL}(\kappa)\;\equiv \;
4\pi\sum_{L''}C^{L''}_{LL'}\;i^{-l+l'-l''}\;\kappa\;
n_{L''}(\kappa,{\bf R}'-{\bf R}),
\end{equation}
and where $C^{L''}_{LL'}$ are the real Gaunt numbers.

For the partial waves explicitly included in the formalism, the 
{\it so called low partial waves} with $l\le l_{low}=2-3$, 
$\alpha_{Rl}(\kappa)$ are the hard sphere phase shifts given by
\[
cot \alpha_{Rl}(\kappa)\;=\;n_l(\kappa a_{Rl})/j_l(\kappa a_{Rl})
\]
and for the remaining $Rl$-chanels, $\alpha_{Rl}(\kappa)$ are the proper
phase shifts. For high $l$'s the latter vanish, and at that point the matrix
to be inverted in (\ref{dyson}) can be truncated.

When the hard sphere radii, $a_R$, are properly chosen and $\kappa^2$
lies below the bottom of the hard sphere continuum, the screened
spherical waves have short range. Therefore, the slope matrix can be calculated
in real space and the method is suitable to treat impurities, defects,
surfaces, etc. It was shown in Ref.\ \cite{andersen94} that the shortest range 
of the screened spherical waves can be achieved for non-overlapping
spheres with $a_R\approx 0.5-0.85\;s^i_R$, depending on the maximal 
orbital quantum number $l$ of the partial waves explicitly included in the 
formalism. The $s^i_R$ denotes 
the inscribed or touching sphere radii. In the KKR community, it is
customary to determine the $\alpha_{Rl}(\kappa)$'s as the phase shifts of
repulsive  potentials.

Because a screened spherical wave has pure
$(l,m)$ character only on its own $a$-sphere, the matching condition in Eq.\ 
(\ref{wavefc}) should be set up at this sphere.
The connection onto the potential sphere (s) is done by introducing a free 
electron solution $\varphi_{Rl}(\epsilon,r_R)\;Y_L({\hat r}_R)$ from the 
potential sphere back to the hard sphere, which joins continuously and 
differentiable to the partial wave, $\phi_{RL}(\epsilon,{\bf r}_R)$, at 
$s_R$ and continuously to the screened spherical wave at $a_{Rl}$. The radial 
part of this backwards extrapolated free-electron solution, after normalizing 
it to one at its $a$-sphere, is given by

\begin{equation}\label{freefi}
\varphi^a_{Rl}(\epsilon,r) \equiv \frac{\varphi_{Rl}(\epsilon,r)}
{\varphi_{Rl}(\epsilon,a_{Rl})} = f^a_{Rl}(\kappa,r)\;+\;
g^a_{Rl}(\kappa,r)\;D^a_{Rl}(\epsilon),
\end{equation}
where $D^a_{Rl}(\epsilon)$ is the logarithmic derivative of 
$\varphi_{Rl}(\epsilon,r)$ calculated at the hard sphere $a_{Rl}$. This can be
determined from the matching condition between $\phi_{RL}(\epsilon,r)$ and 
$\varphi_{RL}(\epsilon,r)$ at $r_R=s_R$,

\begin{equation}\label{da}
D^a_{Rl}(\epsilon)\;\equiv\;{\mathcal D}\{\varphi^a_{Rl}(\epsilon,a_{Rl})\}
\;=\;-\frac{f^a_{Rl}(\kappa,s_R)}{g^a_{Rl}(\kappa,s_R)}\;
\frac{{\mathcal D}\{\phi_{Rl}(\epsilon,s_R)\}-
{\mathcal D}\{f^a_{Rl}(\kappa,s_R)\}}{{\mathcal D}\{\phi_{Rl}(\epsilon,s_R)\}
-{\mathcal D}\{g^a_{Rl}(\kappa,s_R)\}}.
\end{equation}

The relation between the values of the free electron function at $a$ and
the partial wave at $s$ is
\begin{equation}\label{normfunc}
\frac{\varphi_{Rl}(\epsilon,a_{Rl})}{\phi_{Rl}(\epsilon,s_R)}\;=\;
\frac{\varphi_{Rl}(\epsilon,a_{Rl})}{\varphi_{Rl}(\epsilon,s_R)}\;=\;
\frac{1}{f^a_{Rl}(\kappa,s_R)}\;
\frac{{\mathcal D}\{\phi_{Rl}(\epsilon,s_R)\}-
{\mathcal D}\{g^a_{Rl}(\kappa,s_R)\}}{{\mathcal D}\{f^a_{Rl}(\kappa,s_R)\}
-{\mathcal D}\{g^a_{Rl}(\kappa,s_R)\}},
\end{equation}

In Fig.\ 1 we have plotted the logarithmic derivative at $a=0.7\;w$, where
$w$ denotes the average Wigner-Seitz radius, and the 
normalization function $\varphi_{Rl}(\epsilon,a_{Rl})$ given in (\ref{normfunc})
in the case of {\it fcc\/} Ga. The logarithmic 
derivative is a never increasing function of energy and it has a pole
above the top of the respective band. Between these poles $D^a_{Rl}(\epsilon)$
is smooth functions of energy, which varies more slowly than
${\mathcal D}\{\phi_{Rl}(\epsilon,s)\}$, because $a\;<\;s$.
The poles of $D^a_{Rl}(\epsilon)$ depend on the 
representation $(a)$ and they are not related directly to the band structure. 
The $\varphi_{Rl}(\epsilon,a_R)$ from the figure was obtained for 
partial waves normalized in the $w$-sphere. It is always a smooth function 
of the energy and
vanishes at the poles of $D^a_{Rl}(\epsilon)$.

The slope matrix, Eq.\ (\ref{dyson}), the logarithmic derivative, Eq.\
(\ref{da}), and the normalization function, Eq.\ (\ref{normfunc}),
play a central role in the present implementation of the EMTO Theory.

\subsection{Kink cancellation equation}

Using the free electron solutions from (\ref{ssw}) and (\ref{freefi}) and the
partial waves $\phi_{Rl}(\epsilon,{\bf r}_R)$ we can introduce a complete basis
set defined in the whole space. These exact muffin tin orbitals or kinked 
partial waves may be written in the form

\begin{equation}\label{kpw}
\bar{\psi}^a_{RL}(\epsilon,{\bf r}_R) = \left(\phi^a_{Rl}(\epsilon,r_R)\;
-\;\varphi^a_{Rl}(\epsilon,r_R)\right)\;Y_L({\hat r}_R)\;+\;
\psi^a_{RL}(\kappa^2,{\bf r}_R),
\end{equation}
where the radial part of the functions $\phi^a_{Rl}$ and $\varphi^a_{Rl}$
are truncated outside the sphere of radius $s_R$ and outside $s_R$ and 
inside $a_R$ , respectively. Moreover, the $l\le l_{low}$ projection of the 
$\psi^a_{Rl}$ function is truncated inside the sphere of radius $a_R$, 
while the high-$l$ components penetrate into the hard spheres.
The $\bar{\psi}^a_{RL}(\epsilon,{\bf r}_R)$ functions are continuous and
differentiable in the whole space, except at the hard spheres, where they have 
non zero kinks. In Eq.\ (\ref{kpw}) the partial waves are renormalized according
to Eq.\ (\ref{freefi})

\begin{equation}\label{phia}
\phi^a_{RL}(\epsilon,r_R)\;\equiv\;\frac{\phi_{RL}(\epsilon,r_R)}
{\varphi_{RL}(\epsilon,a_R)}.
\end{equation}
From Eq.\ (\ref{normfunc}) and (\ref{phia}) it is immediately seen that the 
multiplicativ normalization of the partial waves does not enter in the
expression of the kinked partial wave.
Forming a linear combination of the kinked partial waves,

\begin{equation}
\Psi_j({\bf r})\;=\;\sum_{RL}\;{\bar\psi}^a_{RL}(\epsilon_j,{\bf r}_R)\;
v^a_{RL,j},
\end{equation}
and asking for the kinks be canceled we arrive to the kink 
cancellation or screened KKR equations

\begin{eqnarray}\label{kkreq}
\sum_{RL}\;K^a_{R'L'RL}(\epsilon_j)\;v^a_{RL,j}\;&\equiv& \nonumber\\
\sum_{RL}\;
a_{R'}\; \Bigl [S^a_{R'L'RL}(\kappa^2_j)\;-\;\delta_{R'R}\delta_{L'L}\;
D^a_{RL}(\epsilon_j)\Bigr]\;v^a_{RL,j}\;&=&\;0
\;\;\mbox{for all}\;\;R'L'.
\end{eqnarray}
Here we have $l,l' \le l_{low}$.
The solutions of this equation are the one-electron energies and 
eigenfunctions, which, using Eq.\ (\ref{wavefc}) are given by

\begin{equation}
u_{RL,j}\;=\;\frac{v^a_{RL,j}}{\varphi_{Rl}(\epsilon_j,a_R)}.
\end{equation}
It is worth to note that in the final expression of the wavefunction 
$\Psi_j({\bf r})$, Eq.\ (\ref{wavefc}), the backwards extrapolated free 
electron solution does not enters.

In the case of translation symmetry in Eq.\ (\ref{kkreq}) $R$ and $R'$ run
over the atoms in the primitive cell only, and the slope matrix, and 
thus the kink matrix $K^a_{R'L'RL}$ as well, depend on the Bloch vector 
${\bf k}$ from the first Brillouin zone.
In Fig.\ 2 we plotted the diagonal elements of the {\it fcc\/} slope matrix
(symbols) calculated at the center of the Brillouin zone as a function of the 
dimensionless energy parameter $(\kappa w)^2$. In this calculation the 
matrix inversion in (\ref{dyson}) was performed in real space for $5$ 
coordination shells plus the central site using the $s,p$ and $d$ orbitals 
and $0.7 w$ for the hard sphere radius.
The figure demonstrates the weak and smooth energy 
dependence of the slope matrix up to the bottom of the continuum, 
$(\kappa w)^2\approx 6$.  Therefore in the practical solution of the 
kink cancellation equation (\ref{kkreq}) the slope matrix can be estimated
using a Taylor expansion around a fixed energy $\kappa_0^2$,

\begin{equation}\label{taylor}
S^a_{R'L'RL}(\kappa^2)\;=\;S^a_{R'L'RL}(\kappa^2_0)+
\frac{1}{1!}{\dot S}^a_{R'L'RL}(\kappa^2_0)(\kappa^2-\kappa^2_0)+...,
\end{equation} 
where the overdot indicates energy derivative. The first and higher order
energy derivatives are calculated analytically as described in Ref.\ 
\cite{tank}. In equation (\ref{taylor}) 
$\kappa^2$ is a complex energy not too far from $\kappa^2_0$. 
In Fig.\ 2 the solid lines were calculated with a fourth order expansion
around $\kappa_0=0$. As one can observe, this expansion gives highly 
accurate energy dependence of the slope matrix over an energy range of 
approximately $(-1,+1) Ry$.

\subsection{The electron density}

In order to construct the new one-electron potential for a self-consistent 
calculation first we need to construct the electron density given by

\begin{equation}\label{n(r)}
n({\bf r})\;=\;\sum_j^{\epsilon_j\le\epsilon_F} |\Psi_j({\bf r})|^2,
\end{equation}
where the sum runs over the one-electron states below the Fermi level
$\epsilon_F$. In the present implementation of the method instead of calculating
explicitly the wave functions (\ref{wavefc}) and performing the summation in
Eq.\ (\ref{n(r)}) we introduce the path operator $g^a_{R'L'RL}(z,{\bf k})$ 
defined for a complex energy $z$ and Bloch vector ${\bf k}$ by

\begin{equation}\label{pathop}
\sum_{R''L''}\;K^a_{R'L'R''L''}(z,{\bf k})\;g^a_{R''L''RL}(z,{\bf k})\;=
\;\delta_{R'R} \delta_{L'L}.
\end{equation}
This function is analytic in the complex plane and it has poles at the 
one-electron energies along the real axis. Therefore, using the residue
theorem, for the total number of electrons we find

\begin{equation}\label{nos}
N(\epsilon_F)\;=\;\frac{1}{2\pi i}\;\oint_{\epsilon_F}\sum_{R'L'RL} 
\int_{BZ}\;g^a_{R'L'RL}
(z,{\bf k})\; {\dot K}^a_{RLR'L'}(z,{\bf k})\;d{\bf k}\;dz,
\end{equation}
where the first integration is performed on a complex contour and the 
second one in the first Brillouin zone. The contour is chosen
in a way that it cuts the real axis below the bottom of the valence 
band and at $\epsilon_F$. In (\ref{nos}) $l,l' \le l_{low}$.
The $z$ dependent partial waves,
$\phi_{Rl}(z,r_R)$, and logarithmic derivatives, $D^a_{Rl}(z)$, are 
obtained by solving Eq.\ (\ref{radsch}) for complex energy. The 
energy derivative of the kink matrix,

\begin{equation}
{\dot K}^a_{R'L'RL}(z,{\bf k})\;=\;a_{R'}\; \Bigl [{\dot S}^a_{R'L'RL}
(z-v_0,{\bf k})\;- \;\delta_{R'R}\delta_{L'L}\;{\dot D}^a_{RL}(z)\Bigr],
\end{equation}
is calculated by taking the derivatives of Eq.\ (\ref{da}) and
(\ref{taylor}), where the energy derivatives of the basis functions 
$\{f^a,g^a\}$ are calculated analytically. The energy derivative of 
the logarithmic derivative function is given by \cite{andersen94}

\begin{equation}
\frac{\partial {\mathcal D}\{\phi_{Rl}(z,s_R)\}}{\partial z}\;=\;
-\frac{\int^{s_R}_0\phi^2_{Rl}(z,r_R)\;r^2_R\;dr_R}{s_R\;
\phi^2_{Rl}(z,s_R)}.
\end{equation}
Because the eigenvectors are normalized as (see Ref.\ \cite{andersen94})

\begin{equation}\label{normalization}
\int \Psi^*_j({\bf r})\;\Psi_j({\bf r})\;d{\bf r}\;=\;
\sum_{R'L'RL}\;v^{a\;*}_{R'L',j}\;{\dot K}^a_{R'L'RL}(\epsilon)
\;v^a_{RL,j}
\end{equation}
the expression (\ref{nos}) gives the exact number of states at the
Fermi level for the muffin tin potential (\ref{mtpot}).
In (\ref{normalization}) the negligible terms due to the overlap between 
$s$-spheres are omitted \cite{third}.

Inside the unit cell at $R$ the electron density in terms of the path operator 
can be expressed as

\begin{equation}\label{n(r)s}
n({\bf r}_R)\;=\;\frac{1}{2\pi i}\oint_{\epsilon_F} \sum_{L'L}
Z^a_{RL'}(z,{\bf r}_R)
\int_{BZ} \tilde g^a_{RL'RL}(z,{\bf k})d{\bf k} Z^a_{RL}(z,{\bf r}_R)dz,
\end{equation}
where we have introduced the functions

\begin{eqnarray}
Z^a_{RL}(z,{\bf r}_R)\;=\;\left\{\begin{array}{lll}
\phi_{Rl}^a(z,r_R)\;Y_L({\hat r}_R)\;&\mbox{if}\;\;l\le\l_{low}\;
\mbox{and}\;r_R\le s_R\\
\varphi^a_{Rl}(z,r_R)\;Y_L({\hat r}_R)\;&\mbox{if}\;\;l\le\l_{low}\;
\mbox{and}\;r_R > s_R\\
-j_l(z,r_R)\;Y_L({\hat r}_R)\;&\mbox{if}\;\;l>l_{low}\;\mbox{for all}
\;r_R\end{array}\right. ,
\end{eqnarray}
and where the sums over $l'$ and $l$ include the high-$l$ terms as well.
These functions are equivalent to the scattering solutions of Faulkner 
and Stocks \cite{faulkner80}. In Eq.\ (\ref{n(r)s}) we have introduced the
following matrix

\begin{eqnarray}
\tilde g^a_{RL'RL}(z,{\bf k})\equiv\left\{\begin{array}{lllll}
g^a_{RL'RL}(z,{\bf k})\;&\mbox{if}\;\;l,l'\le l_{low}\\
\sum_{R''L''}g^a_{RL'R''L''}(z,{\bf k})S^a_{R''L''RL}(z,{\bf k})
\;&\mbox{if}\;\;l'\le l_{low}\;\mbox{and}\;l> l_{low}\\
\sum_{R''L''}S^a_{RL'R''L''}(z,{\bf k})g^a_{R''L''RL}(z,{\bf k})
\;&\mbox{if}\;\;l'> l_{low}\;\mbox{and}\;l\le l_{low}\\
\sum_{R''L''}\sum_{R'''L'''}S^a_{RL'R''L''}(z,{\bf k})\\
\times g^a_{R''L''R'''L'''}(z,{\bf k}) S^a_{R'''L'''RL}(z,{\bf k})
\;&\mbox{if}\;\;l',l> l_{low}\end{array}\right. 
\end{eqnarray}
where the high-low and the low-high subblocks of the slope matrix
are calculated by the usual blowing-up technique \cite{andersen85}.

In principle Eq.\ (\ref{nos}) and (\ref{n(r)s}) give the exact number of
states and electron density. However, in some cases, like for the metals 
from the $II B$ and $III-V A$ groups, where one of the $d$ bands is completely
filled, around the top of this band the normalization function (\ref{normfunc})
goes through zero. This happens, for example, in the case of {\it fcc\/} Ga 
around the top of the $3d$ band, as it can be seen from Fig. \ 1. For this 
energy not only the logarithmic derivative but also its energy derivative 
${\dot D}^a_{Rl}$, 
 appearing in the diagonal of the ${\dot K}^a_{R'L'RL}$ matrix, has poles.
In order to cancel these nonphysical poles we rewrite the expression for the 
number of states as

\begin{eqnarray}\label{nosc}
N(\epsilon_F)\;&=&\;\frac{1}{2\pi i}\;\oint_{\epsilon_F} G(z) dz,
\end{eqnarray}
where

\begin{eqnarray}\label{noscp}
G(z)\equiv \sum_{R'L'RL}\int_{BZ}
g^a_{R'L'RL}(z,{\bf k})\;{\dot K}^a_{RLR'L'}(z,{\bf k})\;d{\bf k}
-\sum_{RL}
\left [\frac{{\dot D}^a_{Rl}(z)} {D^a_{Rl}(z)}- \sum_{\epsilon^D_{Rl}}
\frac{1}{z-\epsilon^D_{Rl}}\right],
\end{eqnarray}
with $l,l'\le l_{low}$, and that of the electron density as

\begin{eqnarray}\label{n(r)sc}
n({\bf r}_R)&=&\frac{1}{2\pi i}\oint_{\epsilon_F} \sum_{L'L}Z^a_{RL'}(z,{\bf r}_R)
\int_{BZ} \tilde g^a_{RL'RL}(z,{\bf k})d{\bf k} 
Z^a_{RL}(z,{\bf r}_R)dz
\nonumber\\
&+&\frac{1}{2\pi i}\oint_{\epsilon_F} \sum_L \frac{{Z^a}^2_{RL}(z,{\bf r}_R)}
{a_R\;D^a_{Rl}(z)} d z 
-\sum_L\sum_{\epsilon^D_{Rl}}\frac{{Z^a}^2_{RL}(\epsilon^D_{Rl},{\bf r}_R)}
{a_R\;{\dot D}^a_{Rl}(\epsilon^D_{Rl})},
\end{eqnarray}
where $\epsilon^D_{Rl}$ are the zeros of the logarithmic derivative function,
$D^a_{Rl}(\epsilon)$. Because the logarithmic derivative is a smooth 
decreasing function of energy $\epsilon^D_{Rl}$'s can be easily determined 
with high accuracy. The second and third terms from the right hand side 
of (\ref{n(r)sc}) are included only for $l\le l_{low}$.
Using the fact that the residuum of the $1/{D^a_{Rl}}$ around 
$\epsilon^D_{Rl}$ is $1/{{\dot D}^a_{Rl}}$, it is easy to show that 
the poles of ${\dot D}^a_{Rl}(z)$ and those of $1/{\varphi_{Rl}(z,a_R)}^2$ 
are canceled out in Eqs. (\ref{noscp}) and (\ref{n(r)sc}).

From the electron density (\ref{n(r)sc}) we determine the overlapping 
muffin tin wells and repeat the iterations until self consistency, of the 
total energy for example, is achieved. 
In the present implementation of the EMTO Theory the solution
of the Poisson's equation involves the SCA for the shape of the Wigner-Seitz 
cell, therefore, the construction of the muffin tin potential will be discussed 
only within this context.

\section{The SCA-EMTO method}

Equations (\ref{kkreq}) and (\ref{nosc}-\ref{n(r)sc}), derived in the previous
section, constitute the basis of the present method. In order to perform a 
self-consistent calculation one constructs the electron density from the
solutions of the kink cancellation equation and calculates the new one-electron
potential. In this section we describe these steps using the SCA for the
shape of the Wigner-Seitz cell.

In the SCA, for solving the Poisson's equation, we substitute the 
Wigner-Seitz cells by spherical cells with volumes equal to the volumes of the
real cells. If $\Omega_R$ denotes the volume of the Wigner-Seitz cell
(Voronoi polyhedron) centered at ${\bf R}$ we have
$\Omega_R=\Omega_{w_R} \equiv \frac{4\pi}{3} w_R^3$, where $w_R$ is the 
atomic sphere radius. Thus within the SCA, like in the conventional ASA, the 
whole space is "covered" by the $\Omega_{w_R}$ spheres.

\subsection{The SCA charge density}

During the self-consistent calculation the Fermi level of a $N$ electron 
system is determined by solving the $N(\epsilon_F) = N$ equation, 
where $N(\epsilon_F)$ is given in (\ref{nosc}). For this $\epsilon_F$ 
the electron density is constructed from Eq. (\ref{n(r)sc}). Due to the
normalization (\ref{normalization}) the so constructed density is exactly
normalized within the unit cell cell but not within the SCA spheres of volumes
$\Omega_{w_R}$. Therefore, in order to solve the Poisson's equation within
the SCA we have to renormalize the total density inside the spheres. In the 
present implementation of the method this is realized by

\begin{equation}\label{asachd}
n^{SCA}({\bf r}_R) = n({\bf r}_R) + a Y_{00}(\hat{\bf r}_R),
\end{equation}
where the site independent $a$ constant is determined from the condition
of the charge neutrality within the whole unit cell

\begin{equation}\label{asachdp}
\sum_R \int_{\Omega_{w_R}} n^{SCA}({\bf r}_R) d{\bf r}_R = \sum_R Z_R.
\end{equation}
Here $Z_R$ denotes the nuclear charge at $R$. The sum runs over the
atoms from the unit cell, and the integrals are performed inside the
SCA spheres. Throughout this section the charge density is normalized within 
the SCA spheres according to (\ref{asachd}) and (\ref{asachdp}), however, for 
the sake of simplicity we neglect the $SCA$ index for the $n^{SCA}({\bf r})$.

\subsection{The SCA muffin tin potential}

The spherical symmetric potentials, $v_R(r_R)$, that enter in Eq.\ 
(\ref{radsch}) have to be chosen in a way that, together with the parameter 
$v_0$, to give the best approximation to the full potential $v({\bf r})$. 
The original idea in Ref.\cite{andersen96} is to minimize the mean of the 
squared deviation 
between the left and the right hand side of Eq.\ (\ref{mtpot}). This leads 
to a set of integral or differential equations for $v_R(r_R)$ and $v_0$. 
In the non-overlapping muffin tins case the equation for $v_R(r_R)$ 
reduce to the well known expression

\begin{equation}\label{feq}
v_R(r_R) = \frac{1}{4 \pi} \int v({\bf r}) d{\hat{\bf r}_R},
\end{equation}
and $v_0$ reduces to the muffin tin zero, i.e. to the average of the full 
potential calculated in the interstitial region,

\[
\Omega^I \equiv \Omega -\sum_R V_R\equiv \Omega - \sum_R\frac{4\pi}{3}s_R^3,
\]
where $\Omega$ is the volume of the region where the approximation 
(\ref{mtpot}) is valid (unit cell), and $V_R$ denotes the volume of the 
potential sphere.

In the overlapping muffin tins case the equation for the $v_0$ can be written 
in the following simple form \cite{andersen96}

\begin{equation}\label{eqg}
\sum_R\frac{4\pi}{\Omega}\int_0^{s_R}\left[v_R(r_R)\;-\;v_0\right]r_R^2 dr\;+
\;v_0 = \frac{1}{\Omega}\int_{\Omega}v({\bf r}) d{\bf r},
\end{equation}
while the equation for $v_R(r_R)$ involves terms coming from the overlapping
region, and which give rise to kinks of $v_R(r_R)$ when $r_R$ touches other
muffin tin spheres. In the present implementation of the method, instead of 
solving the $v_R(r_R)$ equations, we all the time, for non-overlapping
and for overlapping muffin tin wells as well, fix the $v_R(r_R)$ 
functions to the spherical average of the full potential given by (\ref{feq}). 
In this case from Eq.\ (\ref{eqg}) we get the expression for the $v_0$ as

\begin{equation}
v_0=\frac{1}{\Omega-\sum_R V_R}\left[\int_{\Omega} v({\bf r})d{\bf r}-
\sum_R\int_{V_R} v({\bf r})d{\bf r}\right],
\end{equation}
or

\begin{equation}\label{eqgp}
v_0=\frac{\sum_R\left[\int_{\Omega^I_R} v({\bf r})d{\bf r}-
\int_{\Omega^{ov}_R} v({\bf r})d{\bf r}\right]}
{\sum_R\left[\Omega^I_R-\Omega^{ov}_R\right]},
\end{equation}
where $\Omega^I_R$ is the real interstitial within a Wigner-Seitz cell
centered at $R$ with volume $\Omega_R$, and $\Omega^{ov}_R$ is that part 
of the potential sphere that is outside of the cell $\Omega_R$, i.e.

\begin{equation}
\Omega \equiv \sum_R\Omega_R\;\;\mbox{and}\;\;
\Omega^I_R-\Omega^{ov}_R\equiv \Omega_R - V_R.
\end{equation}
Eq.\ (\ref{eqgp}) assumes the knowledge of the full potential $v({\bf r})$ 
in $\Omega^I_R$ and $\Omega^{ov}_R$ regions. However, the time consuming 
calculation of the full potential can be avoided by using the SCA for the 
unit cell. In the non-overlapping SCA case, i.e. $s_R < w_R$, we have

\[
\Omega^I_R - \Omega^{ov}_R = 4 \pi \int_{s_R}^{w_R} r^2_R dr_R,
\]
and

\begin{equation}
\int_{\Omega^I_R} v({\bf r})d{\bf r}-
\int_{\Omega^{ov}_R} v({\bf r})d{\bf r}=
\int_{s_R}^{w_R} \left[\int v({\bf r})d{\hat{\bf r}_R}\right]r^2_R d r_R,
\end{equation}
while for the overlapping SCA case, i.e. $s_R > w_R$, we have

\[
\Omega^I_R -\Omega^{ov}_R=- 4 \pi \int_{w_R}^{s_R} r^2_R dr_R,
\]
and

\begin{equation}
\int_{\Omega^I_R} v({\bf r})d{\bf r}-
\int_{\Omega^{ov}_R} v({\bf r})d{\bf r}=
-\int_{w_R}^{s_R} \left[\int v({\bf r})d{\hat{\bf r}_R}\right] r^2_R d r_R.
\end{equation}
From these equations we get the expression for the parameter $v_0$
valid within the SCA

\begin{equation}\label{mfftnz}
v_0\;=\;\sum_R\;\int^{w_R}_{s_R}r^2_R\;\left[\int v({\bf r})d{\hat{\bf r}_R}
\right]\;dr_R / \sum_R\;W_R
\end{equation}
where $W_R\equiv 4\pi (w_R^3-s^3_R)/3$. Therefore both of the $v_R(r_R)$ 
function and the $v_0$ parameter are given in terms of the
spherical symmetric part of the full potential.

The many-body part, $\mu_{xc}[n({\bf r})]$, of the one-electron effective 
potential,

\begin{equation}
v({\bf r})\;=\;v^C({\bf r})\;+\;\;\mu_{xc}[n({\bf r})],
\end{equation}
is calculated within the local density or generalized gradient approximation, 
while the electrostatic part is derived solving the Poisson's equation,

\begin{equation}
\nabla^2 v^C({\bf r})\;=\;-8\pi\Bigl[n({\bf r})\;-\;
\sum_R\;Z_R\;\delta(r_R)
\Bigr],
\end{equation}
for the electronic and nuclear charge densities. 
The electrostatic potential can be divided into intercell and intercell 
component. The spherical symmetric part of the intercell or Madelung potential 
is given by

\begin{equation}\label{madpot}
v^M_R(r_R)\;=\;\frac{1}{w}\;\sum_{R'L'}\;M_{RLR'L'}\;Q_{R'L'}\;\;
\mbox{with}\;\;L=(0,0),
\end{equation}
where $M_{RLR'L'}$ is the Madelung matrix, which can be evaluated by the
usual Ewald technique, and 

\begin{equation}
Q_{RL}\;=\;\frac{\sqrt{4\pi}}{2l+1}\int_{\Omega_{w_R}}\;\Bigl(\frac{r_R}{w}
\Bigr)^l\;\Bigl[n_R({\bf r}_R)-Z_R\delta(r_R)\Bigr]\;Y_L({\hat r}_R)\;
d{\bf r}_R.
\end{equation}
The Hartree part of the intracell Coulomb potential can be obtained as the 
solution of the Poisson's equation using the proper boundary condition at
the atomic sphere radius. Alternatively, this term is given by

\begin{eqnarray}\label{intra}
v^I_R(r_R)\;=\;\left\{\begin{array}{ll}
8\pi\left [\frac{1}{r_R}\int^{r_R}_0{r'_R}^2n_R(r'_R)d{r'_R}\;+\;
\int^{w_R}_{r_R}r_R'n_R(r'_R)d{r'_R}\right]\;\;\mbox{for}\;\;r_R\le w_R\\
8\pi \frac{1}{r_R}\int^{w_R}_0{r'_R}^2n_R(r'_R)d{r'_R}
\;\;\mbox{for}\;\;r_R > w_R\end{array}\right .,
\end{eqnarray}
that is valid inside the potential sphere $s_R$, for $s_R\ge w_R$ as well as 
for $s_R<w_R$. The total potential within the potential sphere is obtained as 
the sum of Eq.\ (\ref{madpot}), (\ref{intra}), the Coulomb potential of the 
nucleus and the spherical symmetric exchange-correlation potential, namely

\begin{equation}\label{totpot}
v_R(r_R)\;=\;v^M_R\;+\;v^I_R(r_R)\;-\;\frac{2 Z_R}{r_R}\;+\;\mu_{xc R}(r_R).
\end{equation}
If the spherical symmetric part of the exchange-correlation potential is
approximated by $\mu_{xc R}[n_R(r_R)]$ besides the higher order multipole 
moments from (\ref{madpot}), which in many cases can be neglected, all of the 
potential components from (\ref{totpot}) depend only on the spherical 
symmetric density $n_R(r_R)$.

Within the SCA-EMTO method the atomic and potential spheres can be and 
usually they are chosen differently. The sizes of the atomic spheres, $w_R$, 
are fixed by the volume, and the ratio between them should be chosen in a way 
that minimizes the errors coming from approximate solution of the Poisson's 
equation. We have found that the best representation of the potential can 
be achieved by choosing the potential sphere radii, $s_R$, larger or equal 
with the atomic sphere radii. 
For an optimal choice of the potential spheres the 
potentials at $s_R$ should be the same, i.e. $v_R(s_R)\approx constant$ for 
each $R$, and this $constant$ should have the maximum possible value for 
linear overlaps bellow $30-40 \%$.

\section{Applications: test calculations}

In this section we present a few applications of the SCA-EMTO method. We chose
particular systems where the conventional ASA based methods failed and
the inclusion of the correction terms or of the exact potential seemed to 
be unavoidable. First we describe the most important numerical details and 
after we analyze the present results comparing to the available full-potential 
calculations.

\subsection{Numerical details}

The hard sphere radii are chosen at $a_R = 0.7 w$. In the matrix inversion 
from Eq.\ (\ref{dyson}) we includ $79$ sites in the case of $fcc$-based
structures ($fcc, L1_2$ and $L1_0$), and $89$ sites in the case of
$bcc$-based structures ($bcc, B2$ and $B32$). The Taylor 
expansion of the slope matrix is carried out for $\kappa_0 = 0$ and 
includes terms up to the fourth to sixth order energy derivative.

The path operator is calculated for $16-32$ complex energy points,
depending on the band structure,
distributed exponentially on a semi-circular contour. The $k$-point
sampling is performed on a uniform grid in the $3D$ Brillouin zones.
All the calculations are scalar-relativistic and employ the frozen-core 
approximation. The basis set include $s$, $p$ and $d$ orbitals and the 
valence electrons are treated self-consistently within the local
density approximation to density functional theory using the 
Ceperley and Alder \cite{ceperley80} exchange-correlation functional 
and parametrized by Perdew and Wang \cite{perdew92}.
The atomic sphere radii, $w_R$-s, are chosen in a way that the
atomic spheres should have the same volumes as the corresponding 
Voronoi polyhedra. The electrostatic and exchange-correlation contribution 
to the total energy is 
calculated within the SCA as described, for example, in Refs.\ 
\cite{vitos94,skriver84}. The kinetic energy is given by

\begin{equation}
T^{SCA} = \frac{1}{2\pi i}\;\oint_{\epsilon_F} z G(z) dz-
\sum_R \int_0^{w_R} v_R(r_R) n_R(r_R) r_R^2 dr_R,
\end{equation}
where the first term from the right hand side is the sum of the one electron
energies and $G(z)$ is given in (\ref{noscp}).

\subsection{Results}

Before starting on the evaluation of the results we address the question of
the accuracy of the Taylor expansion for the slope matrix, Eq.\ (\ref{taylor}).
In Fig.\ 3 the total energy of {\it fcc\/} Cu is shown for different number 
of terms included in the Taylor expansion. The inclusion of the fourth order 
energy derivative term changes the total energy by $1.13$ mRy. The effect
of the fifth order term is already less than $0.2$ mRy and that of the 
sixth order term is about $0.04$ mRy. Therefore, we conclude that for a
reasonable accuracy it is sufficient to include five terms in the expansion of 
the slope matrix, viz. up to the fourth order energy derivatives.
In the case of open structures, wide energy bands or semicore states, however, 
more terms should be included \cite{vitos}.

The variations of the SCA-EMTO total energy with the potential sphere radius
for {\it fcc\/} and {\it bcc\/} Cu are shown on Fig.\ 4. For this test 
calculation the total potential from (\ref{mfftnz}) was weighted by the 
valence part of the total density according to

\begin{equation}
v_0\;=\;\frac{\sum_R\;\int^{w_R}_{s_R}r^2_R\;\left[\int n({\bf r}) v({\bf r})
d{\hat{\bf r}_R} \right]\;dr_R}{ \sum_R 
\;\int^{w_R}_{s_R}r^2_R\;\left[\int n({\bf r})d{\hat{\bf r}_R} \right]\;dr_R},
\end{equation}
and the cut-off in (\ref{n(r)sc}) for the spherical part of the total density
was $l_{max} = l'_{max} = 8$. The inscribed sphere
radii are $s^i_{fcc}=0.91 w_{fcc}$ and $s^i_{bcc}=0.88 w_{bcc}$, where the 
theoretical atomic sphere radii, $w_{fcc}$ and $w_{bcc}$ are shown on the 
figure. The total energy 
in both of the {\it fcc\/} and {\it bcc\/} structures beginning from 
$s\approx 0.80 w$ becomes almost flat with a negligible slope up to 
$s\approx 1.20 w$, which means about $32-36 \%$ linear overlap for
the {\it fcc\/} and {\it bcc\/} structures, respectively. For $s>0.80 w$
the further increase of the potential sphere radius has little effect on the
energy, that means the potential in the corners of the Wigner-Seitz cell
is, with a very good approximation, constant. However, for big overlaps,
$s>1.20 w$, the errors coming from the overlap region, and neglected
in the kink-cancellation equation and in the charge density as well,
become important \cite{third}.

There is a comprehensive study of the structural stability of the 
transition metals done either by full-potential or by muffin-tin 
or ASA based methods. In the latter case correction terms are needed 
\cite{skriver85} for calculation of the accurate total energies.
The conventional ASA without correction terms gives, for example, with 
about $2$ mRy lower total energy for the Cu in the {\it bcc\/} phase than 
in the {\it fcc\/} phase. This underestimation of the {\it bcc\/} total 
energy is due to the incorrect kinetic energy term,
and the inclusion of the exact Hartree energy would lower even more the 
{\it bcc\/} energy. From a more sophisticated full-potential
method \cite{methfessel} a structural energy difference of
$E_{bcc}-E_{fcc} \approx 0.5 $ mRy was obtained. This number should be
compared with our results of $0.4$ mRy from Fig.\ 4. One should note
that this difference is almost constant for a wide range of linear overlap.

The second example is the binary Li$_x$Al$_{1-x}$ ordered compound in 
different phases. There are two reasons of this choice: i) this system 
is very well studied through accurate full-potential calculations
\cite{guo89,sluiter90}, and ii) most of the experimentally observed interesting 
trends, the contraction of the volume of the Al-based alloys, the asymmetric 
heats of formation with respect to the equiatomic concentration etc.,
can not be reproduced by an ordinary spherically symmetric calculation.
Besides the three different compositions, $x=0.25,0.50,0.75$, we 
consider the pure Al and Li limits in {\it fcc\/} and {\it bcc\/} phases
as well. For $x=0.25$ and $0.75$ the calculations were performed only for 
the $L1_2$ structure, while for $x=0.50$ we considered
three different structures: $L1_0, B2$ and $B32$.

In Fig.\ 5 the charge density contour plots are shown for pure {\it fcc} Al,
and Al$_3$Li in $L1_0$ structure, as calculated from Eq.\ (\ref{n(r)sc}) 
using $s,p,d$ basis set, and the maximum orbital quantum number
$l'$ included in (\ref{ssw}) was $10$. The agreement between these plots and
those from Ref.\ \cite{guo90} is very good.

The calculated
equilibrium Wigner-Seitz radii and the bulk moduli are tabulated in
Table I and plotted in Figs.\ 6 and 7. In the case of pure Al and Li the 
structure energy differences and in the case of compounds the heats of 
formation are included in the Table and plotted in Fig. \ 8.
The heat of formation is defined as

\begin{equation}
\Delta H \equiv\; E_{Li_xAl_{1-x}}-x\;E_{Li} - (1-x)\; E_{Al},
\end{equation}
where all the energies are obtained for the proper equilibrium volume 
and they are expressed per atom. In Figs.\ 5-7 and Table I the full-potential
values from Ref.\ \cite{sluiter90} are also included.

The mean deviations between the present and the full-potential results
for the equilibrium radii, bulk moduli and heats of formations are
$9.8 \%, 7.5 \%$ and $17 \%$, respectively. Taking into account the minor
discrepancies between the numerical details used in the calculations, 
for example the exchange-correlation functional, the way the core electrons 
were treated etc., and the fact that the full-potential methods have their own 
error limits as well, we can conclude that the agreement between the two sets 
of results is very good. One should appreciate how well the trends obtained
in the full-potential calculation are reproduced by the present method.

\section{Conclusions}

We have presented a self-consistent implementation based on the Green's
function technique of the Exact Muffin-Tin Orbitals Theory, developed by 
O.K. Andersen {\it et al}. The accuracy of the present implementation
was tested on different systems, where we have found a good agreement between 
the present results and the results obtained by full-potential techniques.
In order to gain some experience about the efficiency of the present
method we compare the CPU times of a self-consistent calculation 
of the tight-binding ASA-LMTO method \cite{andersen85}, based also on the 
Green's function technique, and that of the SCA-EMTO method. We found that the 
present implementation of the SCA-EMTO method needs with about $3$ times 
larger CPU time than the tight-binding ASA-LMTO method.

Finally we remark that if the radii of the potential spheres are
chosen to be equal with the radii of the atomic spheres, i.e. $w_r=s_R$,
the SCA-EMTO method can be considered as an ASA based Green's function 
technique that involves the so called {\it combined correction} term 
\cite{andersen87}.
It gives exact one electron energies and charge densities for the 
optimized overlapping muffin tin wells. The natural extension of the 
present SCA-EMTO method to compute the total energies from the output total
charge density via the Full Charge Density technique 
\cite{vitos97,kollar99} is in progress.

\acknowledgments
L.V. acknowledges the interesting and helpful discussions with Prof. O. K.
Andersen, the assistance from Drs. C. Arcangeli and R. W. Tank, and the 
hospitality of the Max-Planck Institute from Stuttgart where the first part of 
this work was performed. Thanks are also due to Dr. A. V. Ruban for his 
valuable observations.
The Swedish Natural Science Research Council, the Swedish Foundation for
Strategic Research and Royal Swedish Academy of Sciences are acknowledged 
for financial support. Center for Atomic-scale Materials Physics is sponsored 
by the Danish 
National Research Foundation. Part of this work was supported by the research 
project OTKA 023390 of the Hungarian Scientific Research Fund.

\section{Appendix}

During the self-consistent procedure the Eq.\ (\ref{pathop}), (\ref{nosc}), 
(\ref{n(r)sc}) and (\ref{totpot}) are solved iteratively.
In order to construct the electron density (\ref{n(r)sc}), as the input for
the next iteration, we have to invert the kink matrix
for each complex energy $z$ along the contour and for each
Bloch vector ${\bf k}$ from the Brillouin zone. For a reasonably
high accuracy we need at least a few hundreds of Bloch vectors in the 
irreducible part of the Brillouin zone, therefore this solution means 
the most time-consuming step of the self-consistent procedure.
Here we apply a similar "two-step" scheme introduced in Ref.\ \cite{abricosov97}
in order to reduce the number of time-consuming iterations. Within
this scheme after each iteration an approximate charge self-consistency 
is achieved by solving self-consistently the following equation
written for the ${\bf k}$-integrated path operator

\begin{equation}\label{kdyson}
g^{a}(z)\;=\;\Bigl[1\;+\;g^{a\;0}(z)\;\Bigl(D^{a\;0}(z)\;-\;
D^{a}(z)\Bigr)\Bigr]^{-1}\;g^{a\;0}(z),
\end{equation}
where
where the index $0$ denotes quantities obtained from the previous
iteration, and which are kept fixed during the solution of Eq.\ (\ref{kdyson}).

In the expression for the number of state (\ref{nosc}), we need the 
${\bf k}$-integrated trace of the product between the path operator and 
the energy derivative of the kink matrix, therefore, a similar equation 
to (\ref{kdyson}) has to be established for the quantity

\begin{equation}
G^a_{R'L'RL}(z)\;\equiv\;\int_{BZ}g^a_{R'L'RL}(z,{\bf k})\;
{\dot K}^a_{RLR'L'}(z,{\bf k})\;d{\bf k}.
\end{equation}
Using the definition of the kink matrix after some manipulations we arrive to 
the equation

\begin{equation}\label{Kdyson}
G^a_{R'L'RL}(z)\;=\;g^a_{R'L'RL}(z)\;\left[\frac{G^{a\;0}_{R'L'RL}(z)}
{g^{a\;0}_{R'L'RL}(z)}\;+\;\delta_{R'R}\delta_{L'L}\;a_R\;\Bigl(
{\dot D}^{a\;0}_{Rl}(z)\;-\;{\dot D}^{a}_{Rl}(z)\Bigr)\right].
\end{equation}
It is worth to note that in this expression we do not have matrix 
multiplication. Finally we mention that as soon as the self-consistency is 
achieved, i.e.

\begin{equation}
D^a_{Rl}(z)\;\rightarrow\;D^{a\;0}_{Rl}(z)\;\;\mbox{and}\;\;
{\dot D}^a_{Rl}(z)\;\rightarrow\;{\dot D}^{a\;0}_{Rl}(z),
\end{equation}
both equations, (\ref{kdyson}) and (\ref{Kdyson}), become exact.

\begin{table}
\caption{The theoretical atomic radii, bulk moduli, heats of formation and
{\it fcc}-{\it bcc} structure energy differences of the ordered AlLi compounds,
and pure Al and Li. The present SCA-EMTO results are compared to the
full-potential FLAPW results, and to the experimental data.}
\begin{tabular}{lccccccccc}
&Structure&&Present calculation&Full-potential$^1$&Experimental \\
\hline
$Al_3Li$&$L1_2$&S(Bohr)&2.904&2.935&2.934$^3$\\
        &      &B(GPa)&74.27&70.31&66.0$^4$ \\
        &      &$\Delta H$(mRy)&-8.21&-8.3&     \\
\hline
$AlLi$&$L1_0$&S(Bohr)&2.881&2.917&     \\
        &    &B(GPa)&51.04&50.41&     \\
        &    &$\Delta H$(mRy)&-11.53&-10.25&     \\
\hline
$AlLi$&$B2$&S(Bohr)&2.837&2.876&     \\
        &  &B(GPa)&55.86&42.09&     \\
        &  &$\Delta H$(mRy)&-13.94&-10.45&     \\
\hline
$AlLi$&$B32$&S(Bohr)&2.863&2.910&2.928$^5$\\
        &   &B(GPa)&59.91&57.75&     \\
        &   &$\Delta H$(mRy)&-21.69&-16.60&-18.5$^6$\\
\hline
$AlLi_3$&$L1_2$&S(Bohr)&2.932&2.965&     \\
        &      &B(GPa)&31.45&28.37&     \\
        &      &$\Delta H$(mRy)&-6.71&-5.0&     \\
\hline
\hline
Al&fcc&S(Bohr)&2.920&2.946&2.991$^2$ \\
  &   &B(GPa) &91.76&82.20&72.8$^2$ \\
  &   &$E^{bcc}-E^{fcc}$(mRy)&2.90&4.59& \\
\hline
Al&bcc&S(Bohr)&2.930&2.951&     \\
  &   &B(GPa) &84.46&84.18&     \\
\hline
Li&fcc&S(Bohr)&3.112&3.124& \\
  &   &B(GPa) &15.47&13.64& \\
  &   &$E^{bcc}-E^{fcc}$(mRy)&0.29&0.50& \\
\hline
Li&bcc&S(Bohr)&3.117&3.128&3.237$^2$ \\
  &   &B(GPa) &15.64&15.25&12.6$^2$ \\
\end{tabular}
$^1$ FLAPW calculation, Ref.\cite{sluiter90};
$^2$ Experimental, Ref.\cite{young91};

$^3$ Experimental, Ref.\cite{sanchez84};
$^4$ Experimental, Ref.\cite{mueller86};

$^5$ Experimental, Ref.\cite{brandes83};
$^6$ Experimental, Ref.\cite{barin77}.
\end{table}

\begin{figure}
\caption[1]{The energy dependence of the logarithmic derivative 
$D^a_{Rl}(\epsilon)$ (solid line), and normalization function 
$\varphi_{Rl}(\epsilon,a_R)$ (dashed line) for the {\it fcc\/} Ga.}
\end{figure}

\begin{figure}
\caption[2]{The diagonal elements of the {\it fcc\/} slope matrix in
the ${\bf k}=(0,0,0)$ point from the Brillouin zone versus 
$(\kappa\;w)^2$. The numbers in parenthesis denote the $(l,m)$ 
quantum numbers. The Taylor expansion included terms up to the
$4$th order energy derivative.}
\end{figure}

\begin{figure}
\caption[3]{Self-consistent SCA-EMTO total energy of {\it fcc\/} Cu for 
different
number of terms included in the Taylor expansion for the slope matrix, Eq.\
\protect{(\ref{taylor})}.}
\end{figure}

\begin{figure}
\caption[4]{Total energy versus potential sphere radius, $s$, for the
{\it fcc\/} and {\it bcc\/} Cu. The calculations were done at the 
theoretical equilibrium atomic sphere radii shown in the figure.}
\end{figure}

\begin{figure}
\caption[5]{Charge density contour plots for Al$_3$Li in $L1_2$ structure
and {\it fcc} Al in units of $0.01 \;electrons/Bohr^3$ obtained as the
output of a self-consistent SCA-EMTO calculations.}
\end{figure}

\begin{figure}
\caption[6]{The change of the atomic radii of the ordered AlLi compounds 
relative to the radius of the {\it fcc} Al. 
The open symbols show the results obtained by the full-potential FLAPW 
calculation from Ref.\cite{sluiter90}. The present SCA-EMTO results are 
shown by closed symbols.  The lines connect the results for the 
{\it fcc\/}-based structures.}
\end{figure}

\begin{figure}
\caption[7]{The theoretical bulk moduli for the ordered AlLi compounds.
For the notation see caption of Fig.\ 6.}
\end{figure}

\begin{figure}
\caption[8]{The theoretical heat of formations for the ordered AlLi compounds.
For the pure Al and Li the {\it fcc}-{\it bcc} structure energy difference
is shown.
For the notation see caption of Fig.\ 6.}
\end{figure}

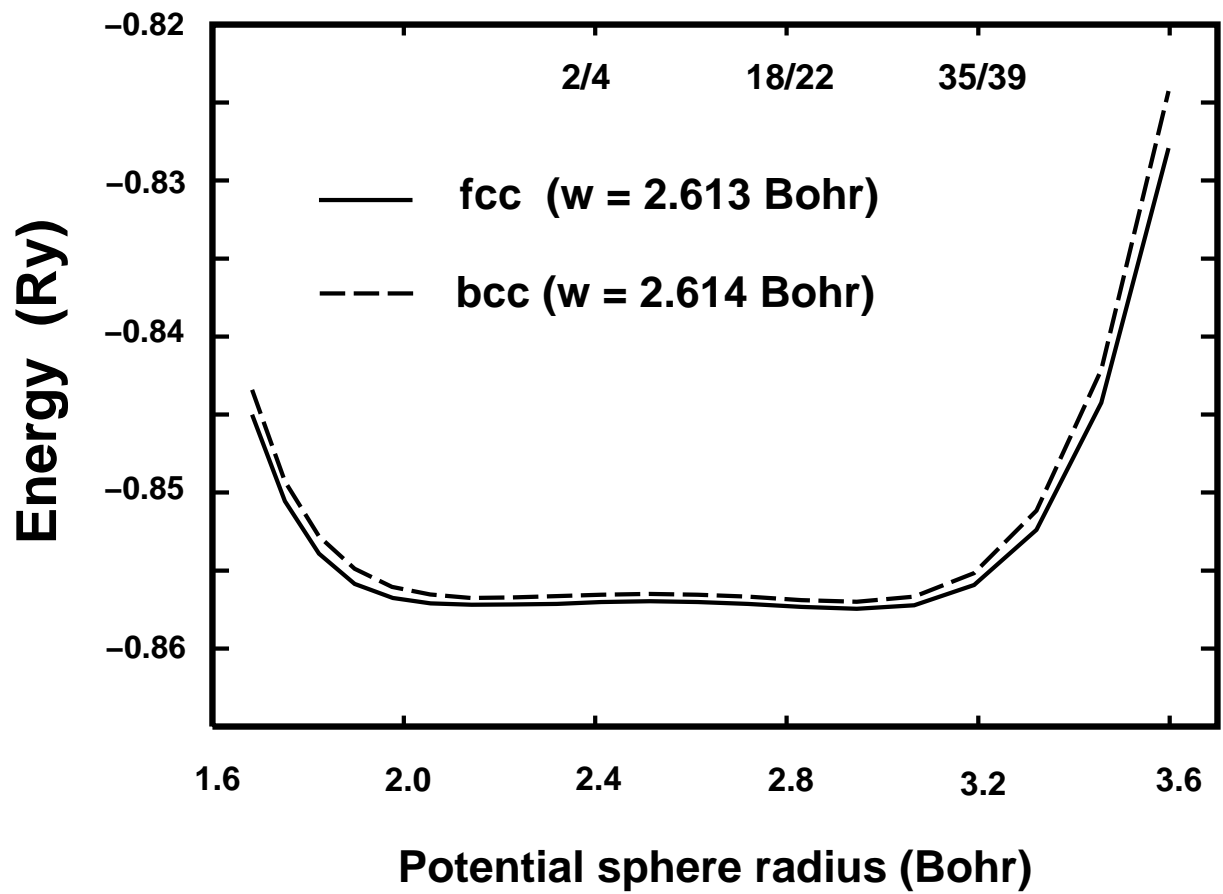
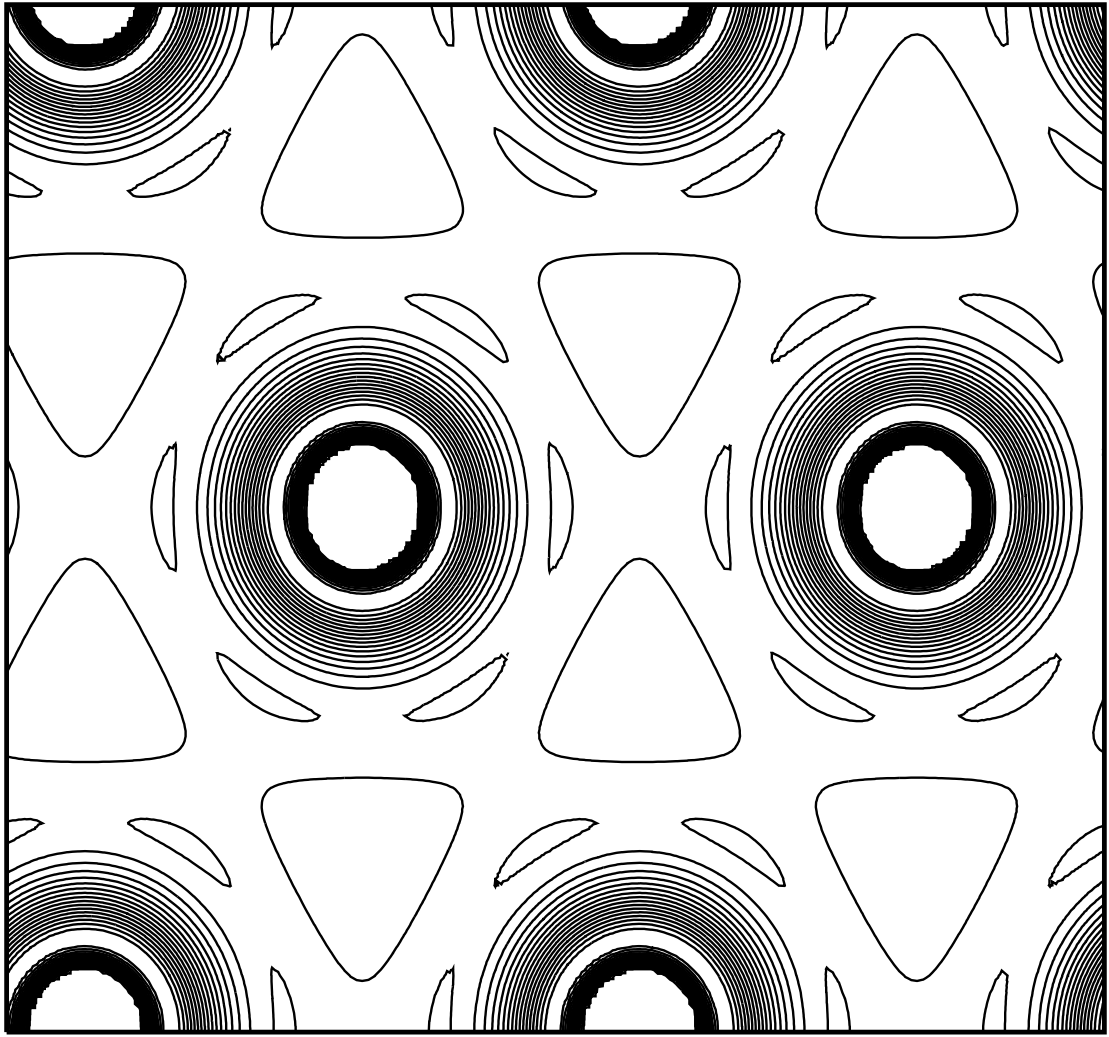
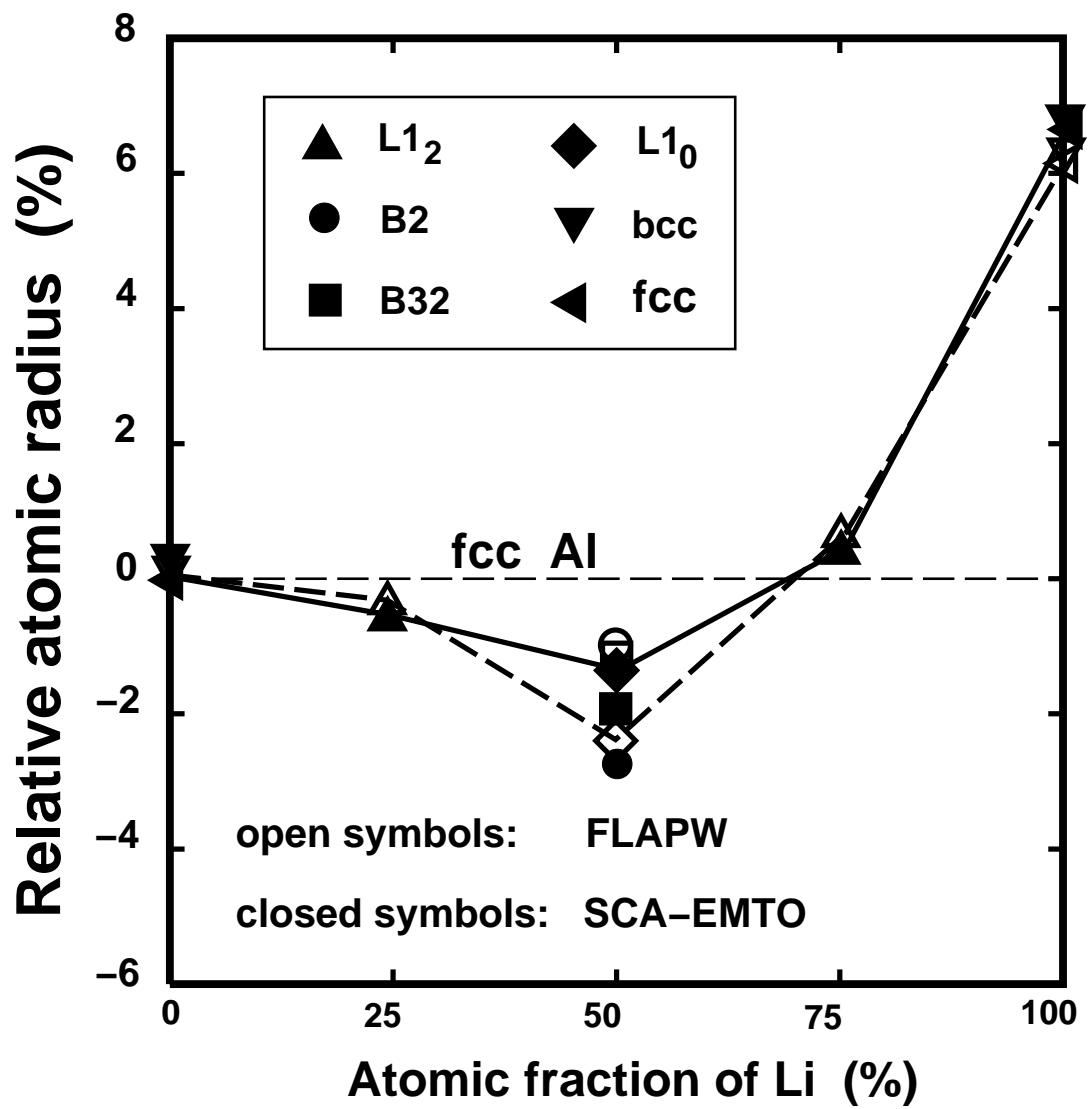
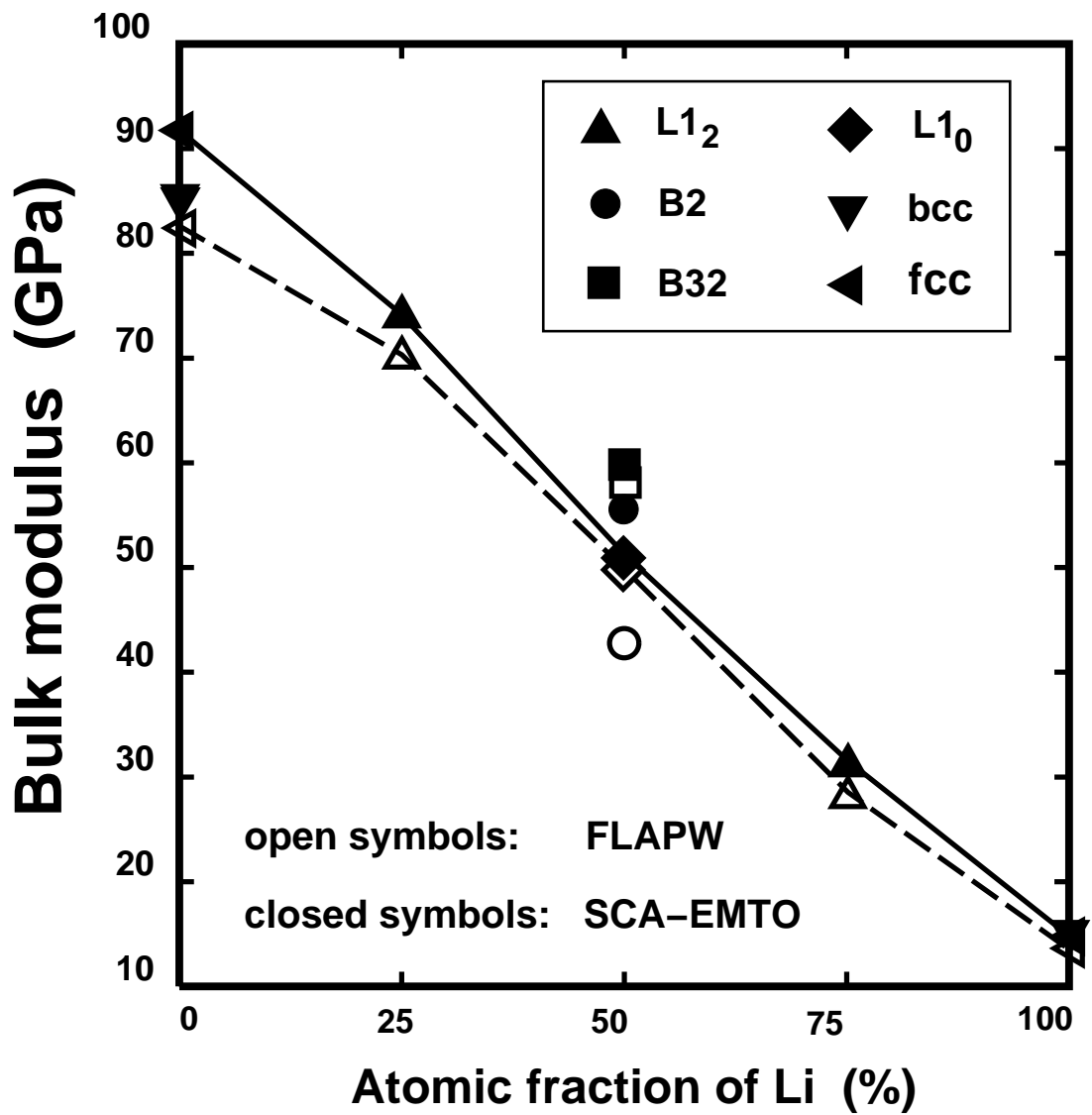
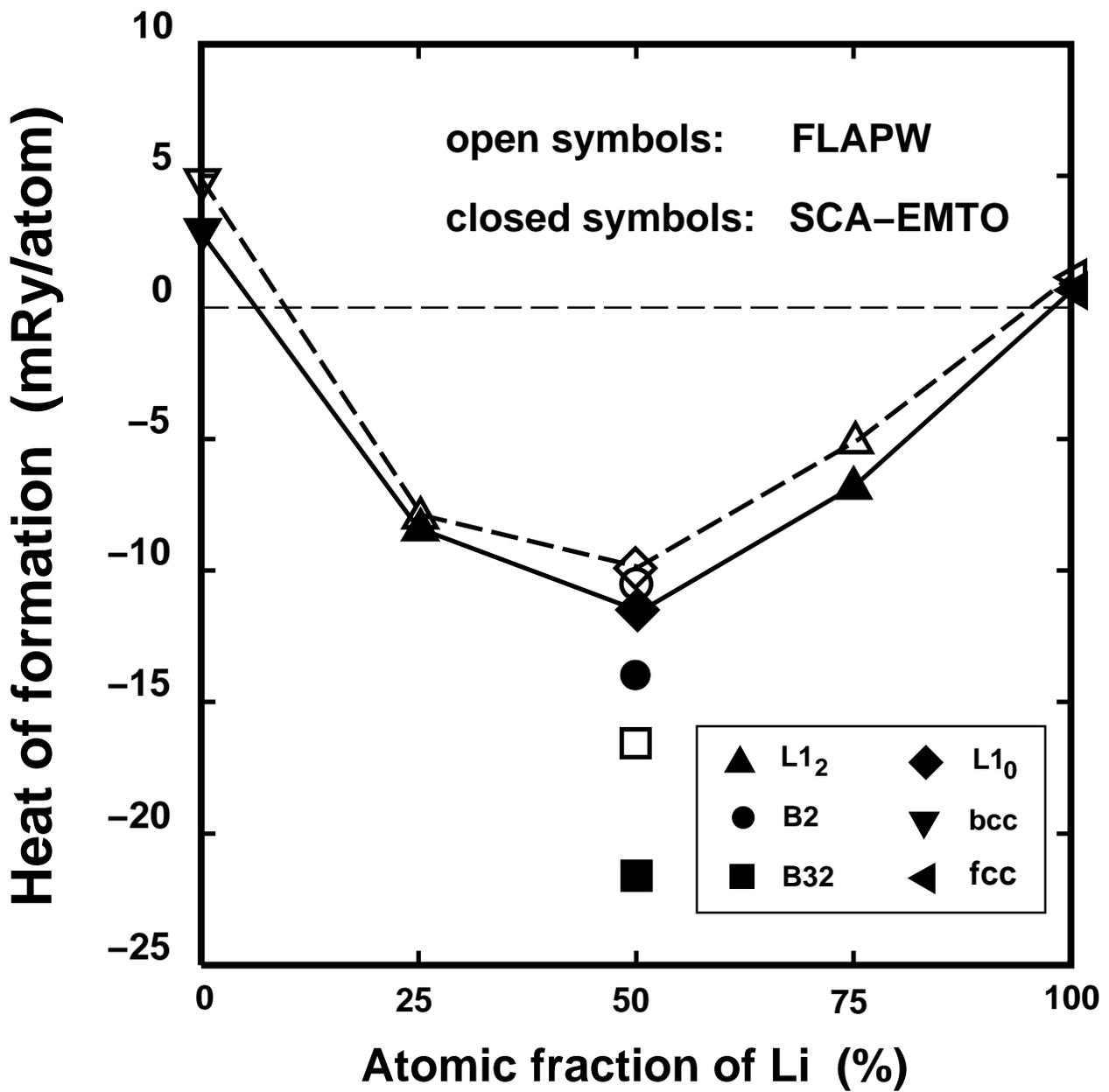
\end{document}